\documentclass[a4paper,reprint,aps,prb,superscriptaddress]{revtex4-1}

\usepackage{amsmath}
\usepackage{txfonts}
\usepackage{mathrsfs}
\usepackage{bm}
\usepackage{braket}
\usepackage{multirow}
\usepackage{graphicx}
\usepackage[para,online,flushleft]{threeparttable}
\usepackage[colorlinks=true,allcolors=blue]{hyperref}
\newcommand{\argmin}{\mathop{\rm arg~min}\limits}

\begin{document}

\title{First-Principles Lattice Dynamics Method for Strongly Anharmonic Crystals}

\author{Terumasa Tadano}
\email{TADANO.Terumasa@nims.go.jp}
\affiliation{International Center for Young Scientists (ICYS), National Institute for Materials Science, Tsukuba 305-0047, Japan}
\affiliation{Research and Services Division of Materials Data and Integrated System (MaDIS), National Institute for Materials Science, Tsukuba 305-0047, Japan}

\author{Shinji Tsuneyuki}
\affiliation{Department of Physics, The University of Tokyo, Tokyo 113-0033, Japan}
\affiliation{Institute for Solid State Physics, The University of Tokyo, Kashiwa 277-8581, Japan}

\begin{abstract}
We review our recent development of a first-principles lattice dynamics method that can treat anharmonic effects nonperturbatively.
The method is based on the self-consistent phonon theory, and temperature-dependent phonon frequencies can be calculated
efficiently by incorporating recent numerical techniques to estimate anharmonic force constants. 
The validity of our approach is demonstrated through applications to cubic strontium titanate, where overall good 
agreement with experimental data is obtained for phonon frequencies and lattice thermal conductivity.
We also show the feasibility of highly accurate calculations based on a hybrid exchange-correlation functional within the
present framework. Our method provides a new way of studying lattice dynamics in severely anharmonic materials where
the standard harmonic approximation and the perturbative approach break down. 
\end{abstract}

\maketitle

\section{Introduction}

Lattice vibration has been one of the most important research subjects in 
the field of condensed matter physics and materials science because of its close
connection with various thermodynamical, dynamical, and transport properties of solids 
including phase stability at finite temperatures, lattice thermal conductivity (LTC), and the
superconducting critical temperature of phonon-mediated superconductors.\cite{ziman}
With the advent of increased computational power and the development of computational
methods based on density functional theory (DFT),\cite{Hohenberg:1964jg,Kohn:1965hf} 
the first-principles calculation of phonons and related properties is widely performed 
in order to interpret experimental results and to design new materials possessing desired physical properties.

While first-principles methods are convenient and reasonably accurate for various applications, 
their validity is always limited by the approximations and assumptions made therein.
In the case of phonon calculations, the harmonic approximation (HA) is usually adopted,
in which only the second derivative of the Born--Oppenheimer (BO) energy surface is considered 
assuming that atomic displacements are sufficiently small compared with interatomic distances.\cite{wallace_book}
The HA is in many cases valid and useful for obtaining phonon dispersion curves 
and discussing phase stability based on the vibrational free energy.~\cite{RevModPhys.73.515}
However, it fails to describe many important properties related to the lattice anharmonicity
such as thermal expansion, LTC, and the temperature and volume dependences of phonon frequencies,
for which we must go beyond the HA.

When the cubic and higher-order anharmonic terms of the BO energy surface are sufficiently small compared with the harmonic one, 
the anharmonic effects can be treated by the many-body perturbation theory (PT).~\cite{PhysRev.128.2589}
The PT has successfully been employed to explain phonon linewidths in
semiconductors\cite{Narasimhan:1991cp,Debernardi:GnfKiRMa} and to understand/predict LTC values of a wide variety of materials.~\cite{2007ApPhL..91w1922B,PhysRevB.82.115427,Cepellotti:2015ke,Jain:2015by,Zeraati:2016in,2011PhRvB..84h5204E,2011PhRvB..84j4302S,Tian2012,Li:2014dd,PhysRevLett.114.095501,Carrete:2014bz,PhysRevB.91.094306,Dekura:2013kv,PhysRevLett.115.205901}
In these calculations, the lowest-order contribution associated with the third-order anharmonic terms is considered
to describe the intrinsic phonon-phonon scattering events, which is of primary importance in a phonon scattering process in semiconductors and insulators.
Despite its great success, the PT is often inadequate and sometimes breaks down completely in important applications.
For example, in quantum crystals such as solid helium\cite{Horner:1967vk} and superconducting sulfur hydride under pressure,\cite{Drozdov:2015hd,PhysRevLett.114.157004}
the zero-point motion is so significant that the anharmonic renormalization of phonon frequencies must be
considered beyond the PT. Furthermore, the PT is not applicable to high-temperature phases of dielectric materials
due to the existence of unstable phonon modes within the HA.
This seriously limits the predictive power of first-principles lattice dynamics methods for emergent thermoelectric 
materials\cite{2011NatMa..10..614D,2014Natur.508..373Z} and hybrid perovskite solar cells.\cite{Kojima:2009el,Frost:2016kl}

To overcome the aforementioned limitations, several DFT-based methods have recently been developed for 
treating anharmonic effects in solids nonperturbatively, 
which rely on the vibrational self-consistent-field theory,\cite{Bowman:1978bu,Monserrat:2013ct}
the self-consistent phonon (SCP) theory,~\cite{Hooton,2008PhRvL.100i5901S,PhysRevB.89.064302,2015PhRvB..92e4301T,vanRoekeghem:2016kd}
or \textit{ab initio} molecular dynamics (AIMD) methods.~\cite{2011PhRvB..84r0301H,Sun:2014bz}
They have been employed to study anharmonic renormalization of phonon frequencies as well as its
effects on the phase stability,\cite{2011PhRvB..84r0301H,Engel:2015bd} 
the superconducting critical temperature,\cite{PhysRevLett.106.165501,Errea:2013jn,PhysRevLett.114.157004,2016PhRvB..93i4525S,Errea:2016js}
and the LTC of high-temperature phases.\cite{2015PhRvB..92e4301T,vanRoekeghem:2016kp}
For example, the self-consistent \textit{ab initio} lattice dynamics (SCAILD)\cite{2008PhRvL.100i5901S} and 
the stochastic self-consistent harmonic approximation (SSCHA)\cite{PhysRevB.89.064302} are SCP-based approaches which
can incorporate the effect of lattice anharmonicity at the mean-field level.
In these methods, anharmonic phonon frequencies, or equivalently, \textit{effective} harmonic force constants are obtained
self-consistently by repeatedly calculating atomic forces in supercells with suitably chosen atomic configurations.
These stochastic algorithms are quite useful since a calculation of fourth-order anharmonic force constants is unnecessary.
More recently, another efficient implementation of the SCP theory was developed,\cite{2015PhRvB..92e4301T} which employs 
anharmonic force constants calculated using the compressive sensing lattice dynamics method.\cite{PhysRevLett.113.185501}
The temperature-dependent effective potential (TDEP) method\cite{2011PhRvB..84r0301H,Hellman:2013ef} is an AIMD-based approach,
in which \textit{effective} harmonic and cubic force constants are extracted from displacement-force data sets
calculated for atomic configurations sampled by AIMD. 
Although the AIMD-based approaches are not valid in the low-temperature region
due to their inability to account for the zero-point motion, they should be accurate in the high-temperature region because 
anharmonic effects are fully included.
As a result of these continuous efforts, an accurate first-principles modeling of lattice vibration is becoming practical 
even for severely anharmonic materials.

In this paper, we review the recent developments of first-principles lattice dynamics methods for strongly anharmonic materials,
particularly focusing on one of the SCP-based approaches developed by the authors.\cite{2015PhRvB..92e4301T}
Using microscopic force constants estimated by supercell methods as inputs,\cite{Esfarjani2008,Tadano2014,PhysRevLett.113.185501} our method can calculate temperature-dependent phonon frequencies and LTC efficiently.
We demonstrate the validity of our method by applying it to cubic SrTiO$_{3}$, which is realized only at temperatures above 105 K.\cite{doi:10.1143/JPSJ.26.396}
In Sect.~\ref{sec:method}, we present theoretical formulations and some numerical techniques that underlie our SCP approach.
In Sect.~\ref{sec:appl}, we show the validity of the approach by carefully comparing our numerical results with available experimental data. In addition, the possibility of highly accurate calculation based on a hybrid exchange-correlation functional
is demonstrated. Finally, we make concluding remarks in Sect.~\ref{sec:conclusion}.

\section{Methodology}
\label{sec:method}
\subsection{Taylor expansion potential}
Let us start by expressing the potential energy of an interacting nuclear system $U$ as a Taylor expansion with respect to the
atomic displacement:
\begin{align}
  &U - U_{0} = U_{2} + U_{3} + U_{4} + \dots, \label{eq:Taylor}\\
  &U_{n} = \frac{1}{n!}\sum_{\{\ell,\kappa,\mu\}}\Phi_{\mu_{1}\dots\mu_{n}}(\ell_{1}\kappa_{1};\dots;\ell_{n}\kappa_{n}) u_{\mu_{1}}(\ell_{1}\kappa_{1})\cdots u_{\mu_{n}}(\ell_{n}\kappa_{n}). \label{eq:Un}
\end{align}
Here, $u_{\mu}(\ell\kappa)$ is the displacement of atom $\kappa$ in the $\ell$th unit cell along 
the $\mu~(=x,y,z)$ direction and $\Phi_{\mu_{1}\dots\mu_{n}}(\ell_{1}\kappa_{1};\dots;\ell_{n}\kappa_{n})$ is 
the $n$th-order derivative of the potential energy with respect to displacement, 
which is usually termed the \textit{interatomic force constant} (IFC).
In Eq.~(\ref{eq:Un}), the IFCs are determined at the ground state and therefore have no temperature dependence.

In the HA, all of the anharmonic terms $U_{n} (n>2)$ are neglected.
Then, the Hamiltonian of the system $H_{0}=T+U_{2}$ can be represented as $H_{0}= \frac{1}{2}\sum_{\bm{q}j}\hbar\omega_{\bm{q}j}A_{\bm{q}j}A_{\bm{q}j}^{\dagger}$ with harmonic phonon frequency $\omega_{\bm{q}j}$ and associated displacement operator $A_{\bm{q}j}=b_{\bm{q}j}+b_{-\bm{q}j}^{\dagger}$, 
where $b_{\bm{q}j}$ and $b_{\bm{q}j}^{\dagger}$ are the annihilation and creation operators of a phonon with crystal momentum $\bm{q}$ and branch index $j$, respectively. 
The phonon frequency $\omega_{\bm{q}j}$ can be obtained from the dynamical matrix
\begin{equation}
	D_{\mu\nu}(\kappa\kappa';\bm{q}) = \frac{1}{\sqrt{M_{\kappa}M_{\kappa'}}}\sum_{\ell'}\Phi_{\mu\nu}(0\kappa;\ell'\kappa')e^{i\bm{q}\cdot\bm{r}(\ell')},
\end{equation}
where $M_{\kappa}$ is the mass of atom $\kappa$ and $\bm{r}(\ell)$ is the translational vector of the $\ell$th unit cell.
By diagonalizing the matrix $D(\bm{q})$, one obtains the squared harmonic frequency and the polarization vector as $D(\bm{q})\bm{e}_{\bm{q}j}$ $=\omega_{\bm{q}j}^{2}\bm{e}_{\bm{q}j}$.
Since the harmonic IFCs are temperature-independent, the intrinsic temperature dependence of 
phonon frequencies and polarization vectors $\{\omega_{\bm{q}j}, \bm{e}_{\bm{q}j}\}$ is absent within the HA.
The atomic displacement $u_{\mu}(\ell\kappa)$ can be represented in terms of $A_{\bm{q}j}$ as follows:
\begin{equation}
	u_{\mu}(\ell\kappa) = \frac{1}{\sqrt{NM_{\kappa}}}\sum_{\bm{q}j}\sqrt{\frac{\hbar}{2\omega_{\bm{q}j}}}A_{\bm{q}j}e_{\mu}(\kappa;\bm{q}j) \exp{[i\bm{q}\cdot\bm{r}(\ell)]}. \label{eq:u_in_A}
\end{equation}
Here, $N$ is the number of $\bm{q}$ points in the first Brillouin zone (BZ) and $e_{\mu}(\kappa;\bm{q}j)$ is a component of $\bm{e}_{\bm{q}j}$. By substituting Eq.~(\ref{eq:u_in_A}) into Eq.~(\ref{eq:Un}), we obtain
\begin{equation}
	U_{n} = \frac{1}{n!}\sum_{\{\bm{q},j\}}\Delta(\bm{q}_{1}+\cdots+\bm{q}_{n})V(q_{1};\dots;q_{n})A_{q_{1}}\cdots A_{q_{n}}.\label{eq:Un_in_A}
\end{equation}
Here and in the following, we use $q$ for the shorthand notation of $(\bm{q},j)$, satisfying
$q=(\bm{q},j)$ and $-q=(-\bm{q},j)$. The term $\Delta(\bm{q})$ is 1 if $\bm{q}$ is a vector of the reciprocal lattice $\bm{G}$ and 0 otherwise. The coefficient $V(q_{1};\dots;q_{n})$ is defined as
\begin{align}
    &V(q_{1};\dots;q_{n})= \bigg( \frac{\hbar}{2}\bigg)^{\frac{n}{2}}\frac{\Phi(q_{1};\dots;q_{n})}{\sqrt{\omega_{q_{1}}\cdots\omega_{q_{n}}}}, \label{eq:Vn}\\
    &\Phi(q_{1};\dots;q_{n}) \notag \\
    &= N^{1-\frac{n}{2}}\sum_{\{\kappa,\mu\}}(M_{\kappa_{1}}\cdots M_{\kappa_{n}})^{-\frac{1}{2}} e_{\mu_{1}}(\kappa_{1};q_{1})\cdots e_{\mu_{n}}(\kappa_{n};q_{n}) \notag \\
    & \hspace{3mm} \times \sum_{\ell_{2},\dots,\ell_{n}}\Phi_{\mu_{1}\dots\mu_{n}}(0\kappa_{1};\ell_{2}\kappa_{2};\dots;\ell_{n}\kappa_{n}) e^{i(\bm{q}_{2}\cdot \bm{r}(\ell_{2})+\cdots+\bm{q}_{n}\cdot \bm{r}(\ell_{n}))}. \label{eq:Phi_recip}
\end{align}
Equation (\ref{eq:Un_in_A}) is the $n$th-order potential energy represented in terms of the harmonic displacement operator $A_{q}$.

\subsection{Anharmonic perturbation theory}
\label{sec:MBPT}

To describe the intrinsic phonon scattering processes and the temperature dependence of phonon frequencies, 
we need to consider the anharmonic terms. 
When the anharmonic terms are sufficiently small compared with the harmonic one, 
we can treat them as a perturbation $H'$ of the non-interacting Hamiltonian $H_{0}$ as
\begin{equation}
	H = H_{0} + H' \approx H_{0} + U_{3} + U_{4}. \label{eq:Hamiltonian}
\end{equation}
Here, we omitted fifth- and higher-order terms since their contributions are much smaller than
the cubic and quartic terms. 
Let $\bm{G}_{q}(\omega)$ be the one-phonon Green's function and $\bm{G}_{q}^{0}(\omega)$ be that of the non-interacting system $H_{0}$. 
Then, the following Dyson equation holds:
\begin{equation}
  [\bm{G}_{q}(\omega)]^{-1}=[\bm{G}_{q}^{0}(\omega)]^{-1}-\bm{\Sigma}_{q}(\omega),
\end{equation}
with $\bm{\Sigma}_{q}(\omega)$ being the anharmonic self-energy, which can be estimated within a systematic diagrammatic approximation.
If the self-energy correction is small and the condition $\hbar\omega_{q} \gg |\Sigma_{q}(\omega_q)|$ is well satisfied,
the phonon quasiparticle picture is still valid and the phonon frequency shift $\Delta_{q}$ and the linewidth $\Gamma_{q}$ are
given as $\Delta_{q}=-\frac{1}{\hbar}\mathrm{Re}\Sigma_{q}(\omega_{q})$ and $\Gamma_{q}=\frac{1}{\hbar}\mathrm{Im}\Sigma_{q}(\omega_{q})$.
\begin{figure}
\centering
\includegraphics[width=8.5cm,clip]{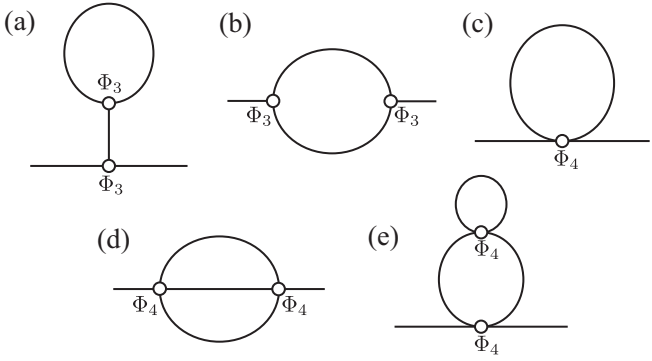}
\caption{Feynman diagrams of anharmonic self-energies up to the second order.
The solid lines represent the free phonon propagator and the small open circles are vertices involving
three or four phonons.
Diagrams (a), (b), and (c) are termed the \textit{tadpole}, \textit{bubble}, and \textit{loop} diagrams, respectively.
Diagrams (a), (c), and (e) are frequency-independent and real, whereas diagrams (b) and (d) are complex
and have frequency dependence.}
\label{fig:diagrams}
\end{figure}

Figure \ref{fig:diagrams} shows the first- and second-order self-energy diagrams
associated with the cubic and quartic anharmonic terms.
The tadpole (a) and bubble (b) are the second-order diagrams resulting from the
cubic terms and their explicit formulae are given as follows:\cite{PhysRev.128.2589}
\begin{align}
&\Sigma_{q}^{\mathrm{tadpole}} = -\frac{1}{\hbar}\sum_{q_{2},j_{1}=\mathrm{TO}}V(-q;q;\bm{0}j_{1})V(\bm{0}j_{1};q_{2};-q_{2}) \frac{2n_{2}+1}{\omega_{\bm{0}j_{1}}}, \label{eq:tadpole} \\
&\Sigma_{q}^{\mathrm{bubble}}(\omega)= \frac{1}{2\hbar} \sideset{}{'}\sum_{q_{1},q_{2},s=\pm1} |V(-q;q_{1};q_{2})|^{2}\notag \\
&\hspace{10mm} \times \left[ \frac{(n_{1} + n_{2} + 1)}{s\omega_{c} + \omega_{q_1} + \omega_{q_2}} 
 - \frac{(n_{1} - n_{2})}{s\omega_{c} + \omega_{q_1} - \omega_{q_2}} \right].
\label{eq:bubble}
\end{align}
Here, $n_{i} = n(\omega_{q_{i}}) = 1/(e^{\beta\hbar\omega_{q_{i}}}-1)$ is the Bose-Einstein distribution 
function and $\omega_{c} = \omega + i0^{+}$ with $0^{+}$ being a positive infinitesimal. 
In addition, the summation in Eq.~(\ref{eq:bubble}) is restricted to the pairs $(q_{1},q_{2})$ satisfying 
the momentum conservation $\bm{q}_{1}+\bm{q}_{2} = \bm{q}+\bm{G}$.

The tadpole diagram is real and therefore gives rise to only a frequency shift.
In Eq.~(\ref{eq:tadpole}), we neglect the processes involving zone-center acoustic modes as the intermediate state $\bm{0}j_{1}$ 
because of the singularity of the matrix element $V(-q;q;\bm{0}j_{1})$ for acoustic modes.
Therefore, the tadpole diagram only accounts for the frequency shift due to the relaxation of internal coordinates 
driven by the cubic anharmonicity. 
The neglected contributions involving acoustic modes can be treated separately by the quasi-harmonic approximation.\cite{Paulatto:2015bb}
The bubble diagram provides the dominant contribution to the phonon linewidth $\Gamma_{q}^{\mathrm{bubble}}(\omega_{q})$, and
the quasiparticle lifetime $\tau_{q}$ can be estimated as $\tau_{q} = 1/ (2\Gamma_{q}^{\mathrm{bubble}}(\omega_{q}))$.

The loop diagram (c) is the first-order contribution from the quartic terms, whose expression is
\begin{equation}
 	\Sigma_{q}^{\mathrm{loop}} = -\sum_{q_{1}} V(q;-q;q_{1};-q_{1}) \frac{2n_1+1}{2}.
 \end{equation}
Since $\Sigma_{q}^{\mathrm{loop}}$ is real, the loop diagram only gives a frequency shift.
The expressions for the higher-order diagrams (d) and (e) associated with the quartic terms are not shown in this paper,
but they can be derived straightforwardly using the Feynman rule.\cite{Tripathi1974,1978PhRvB..18.5859M,1992PhRvB..45.2113P}

First-principles calculations of the bubble and tadpole diagrams have been performed to explain the temperature dependences of the Raman shift and linewidth of group IV and III-IV semiconductors,\cite{Narasimhan:1991cp,Debernardi:GnfKiRMa}.
Recently, there have been many DFT-based estimations of the bubble diagram mostly for the thermal conductivity prediction described in Sect.~\ref{sec:LTC}.
On the other hand, the calculation of the loop diagram has only been reported for relatively simple systems until very recently.\cite{PhysRevB.59.6182,Lazzeri:2003ju,PhysRevLett.99.176802}
This is mainly because the calculation of $\Sigma_{q}^{\mathrm{loop}}$ relies on quartic IFCs, which are technically
more difficult to obtain from DFT than harmonic and cubic IFCs.\cite{2010PhRvB..82j4504R,Errea:2016js}
In Sect.~\ref{sec:IFC}, we will introduce some recent technical developments for estimating the quartic terms.

\subsection{Self-consistent phonon theory}
\label{sec:SCP}

When the anharmonic terms are comparable with the harmonic term, they need to be treated nonperturbatively.
The SCP theory, originally developed by Hooton\cite{Hooton} and followed by more elegant formulations
based on the many-body theory,\cite{1966PhRvL..17...89K,Horner:1967vk,PhysRevB.1.572} is a nonperturbative 
approaches that can treat the anharmonic renormalization of phonon frequencies. 
To derive the SCP equation, we rewrite the Hamiltonian [Eq.~(\ref{eq:Hamiltonian})] as 
\begin{equation}
  H=\mathscr{H}_{0}+(H_{0}-\mathscr{H}_{0} + U_{3}+U_{4})=\mathscr{H}_{0}+\mathscr{H}'.
\end{equation}
Here, $\mathscr{H}_{0}$ is the \textit{effective} harmonic Hamiltonian, which can be written as
$\mathscr{H}_{0}= \frac{1}{2}\sum_{q}\hbar\Omega_{q}\mathcal{A}_{q}\mathcal{A}_{q}^{\dagger}$
with the renormalized phonon frequency $\Omega_{q}$ and the associated displacement operator $\mathcal{A}_{q}$.
Then, the free energy of the original system $F$ can be written as the cumulant expansion
\begin{equation}
  F = \mathscr{F}_{0} + \braket{\mathscr{H}'}_{\mathscr{H}_{0},\mathrm{c}} 
  + \frac{1}{2} \braket{(\mathscr{H}')^{2}}_{\mathscr{H}_{0},\mathrm{c}} +\cdots. \label{eq:F_cumulant}
\end{equation}
Here, $\mathscr{F}_{0} = -\frac{1}{\beta}\log{Z}$ and $\braket{X}_{\mathscr{H}_{0}} = Z^{-1}\mathrm{Tr}{(Xe^{-\beta\mathscr{H}_{0}})}$ with 
the partition function $Z=\mathrm{Tr}{(e^{-\beta\mathscr{H}_{0}})}$, and $\braket{X^{n}}_{\mathrm{c}}$ indicates 
the $n$th-order cumulant of the operator $X$.
The first-order SCP theory only considers the first cumulant in Eq.~(\ref{eq:F_cumulant}), in which the following
Gibbs-Bogoliubov inequality is satisfied:
\begin{equation}
  F \leq \mathscr{F}_{0} + \braket{\mathscr{H}'}_{\mathscr{H}_{0},\mathrm{c}}
  =  \mathscr{F}_{0} + \braket{H-\mathscr{H}_{0}}_{\mathscr{H}_{0}}.
\end{equation}
Since the variational principle holds in the first-order SCP theory, the solution can be
found by minimizing the right-hand side of the above equation with respect to
the adjustable parameters such as anharmonic frequencies $\{\Omega_{q}\}$, polarization vectors, and 
internal coordinates. 
For brevity of the derivation, let us suppose that the polarization vectors and internal coordinates are not altered by
anharmonic effects. Then, from the condition $\partial F_{1}/\partial \Omega_{q} = 0$ with 
$F_{1} =  \mathscr{F}_{0} + \braket{H-\mathscr{H}_{0}}_{\mathscr{H}_{0}}$, the following SCP equation can be readily derived:
\begin{align}
  &\Omega_{q}^{2} = \omega_{q}^{2} + 2\Omega_{q}I_{q}, \label{eq:SCP1}\\
  &I_{q} = \sum_{q_{1}}\frac{\hbar \Phi(q;-q;q_{1};-q_{1})}{4\Omega_{q}\Omega_{q_{1}}}\frac{[2n(\Omega_{q_1})+1]}{2}. \label{eq:SCP2}
\end{align}
By solving Eqs.~(\ref{eq:SCP1}) and (\ref{eq:SCP2}) self-consistently, anharmonic phonon frequencies $\{\Omega_{q}\}$ can be obtained.
The equation is valid even if some of the harmonic phonon frequencies are unstable ($\omega_{q}^{2} < 0$). 
These unstable modes are renormalized by the second term $2\Omega_{q}I_{q}$ to give a stable phonon mode ($\Omega_{q}^{2} \geq 0$). 
Since the formulation assumes the existence of a well-defined Hamiltonian $\mathscr{H}_{0}$, 
an imaginary frequency ($\Omega_{q}^{2} < 0$) is not allowed as a solution of the equation. 

\begin{figure}
\centering
\includegraphics[width=8.5cm,clip]{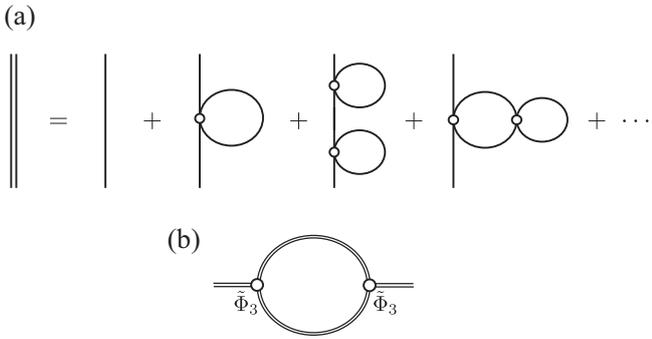}
\caption{(a) Diagrammatic representation of the SCP propagator where the loop diagram is considered. 
The double line represents the SCP propagator and the solid line is that of a free phonon.
(b) Bubble diagram defined on the basis of the SCP propagator.}
\label{fig:SCP}
\end{figure}

The same result has also been obtained on the basis of the Green's function approach.~\cite{doi:10.1021/jp4008834,Hermes:2014jx,2015PhRvB..92e4301T}
Diagrammatically, the SCP theory corresponds to the inclusion of the infinite set of diagrams that can be generated from the loop diagram shown in Fig.~\ref{fig:SCP}(a). 
Therefore, the SCP theory automatically includes the contribution from the higher-order diagram shown in Fig.~\ref{fig:diagrams}(e).
Substituting $\omega_{q}$ for $\Omega_{q}$ in the second term of Eq.~(\ref{eq:SCP1}) gives the correct perturbation limit,
where we have $\Omega_{q} = (\omega_{q}^{2}-2\omega_{q}\Sigma_{q}^{\mathrm{loop}}/\hbar)^{\frac{1}{2}}\approx\omega_{q}-\frac{1}{\hbar}\Sigma_{q}^{\mathrm{loop}}=\omega_{q}+\Delta_{q}^{\mathrm{loop}}$.

In the above derivation of Eqs.~(\ref{eq:SCP1}) and (\ref{eq:SCP2}), we assumed that the polarization vectors are changed by anharmonic effects. Such an assumption is reasonable for simple systems containing a few atoms in the primitive cell\cite{PhysRevLett.106.165501}
but not valid for more complex structures because the polarization mixing (PM) can be significant 
between phonon modes belonging to the same irreducible representation. 
To correctly account for the PM, we extended the SCP equation to include the off-diagonal elements of the loop self-energy and developed an efficient algorithm for solving the equation.\cite{2015PhRvB..92e4301T}
We applied the extended method to cubic SrTiO$_{3}$ and demonstrated the importance of the PM for the soft modes
at high-symmetry $\bm{q}$ points.\cite{2015PhRvB..92e4301T}
The same implementation was also employed to study anharmonic effects in the superconducting hydrogen sulfides.\cite{2016PhRvB..93i4525S}
Moreover, it is possible to extend the theory to include the tadpole diagram.\cite{doi:10.1021/jp4008834}

The deterministic implementation based on the reciprocal formalism [Eqs.~(\ref{eq:SCP1}) and (\ref{eq:SCP2})] is efficient and
applicable to relatively complex structures. 
Also, we can easily evaluate the finite-size effect by increasing the $\bm{q}_{1}$ mesh in Eq.~(\ref{eq:SCP2}) based on the Fourier interpolation, 
which can be significant near a structural phase transition.\cite{Cowley1996585,2015PhRvB..92e4301T}
However, as already mentioned in Sect.~\ref{sec:MBPT}, first-principles calculation of the quartic IFCs necessary 
for $I_{q}$ has been a technical challenge. 
To avoid the cumbersome estimation of the quartic IFCs, several stochastic implementations have been developed\cite{2008PhRvL.100i5901S,PhysRevB.89.064302,Errea:2016js,vanRoekeghem:2016kd} and applied to study anharmonic effects in
simple metals,\cite{2008PhRvL.100i5901S,Lian:2015gk} light-element superconductors,\cite{Errea:2013jn,PhysRevLett.114.157004}
transition metal dichalcogenides,\cite{Leroux:2015dp} and simple perovskite materials.\cite{vanRoekeghem:2016kp}
In these approaches, \textit{effective} harmonic IFCs renormalized by anharmonic effects are estimated by randomly displacing
atoms in a supercell. The stochastic methods should be more accurate than the present deterministic approach [Eqs.~(\ref{eq:SCP1}) and (\ref{eq:SCP2})] because they do not require the expansion of Eq.~(\ref{eq:Taylor}) nor the truncation of higher-order terms carried out
in Eq.~(\ref{eq:Hamiltonian}). However, since a large number of DFT calculations is generally needed to obtain convergence,
the stochastic approaches are more computationally intensive than the deterministic one.

\subsection{Lattice thermal conductivity}
\label{sec:LTC}

The LTC has been intensively studied as it plays an important role in 
optimizing the thermoelectric figure of merit $ZT$. 
Most of the theoretical predictions of LTC rely on either the Boltzmann transport equation (BTE)
or the MD method. 
According to the formulation of BTE for phonons developed by Peierls,\cite{Peierls} the LTC is given as
\begin{equation}
  \kappa = \frac{\hbar}{NVk_{\mathrm{B}}T} \sum_{q}\omega_{q}\bm{v}_{q} \otimes\bm{f}_{q}n_{q}(n_{q}+1),
  \label{eq:LTC}
\end{equation}
where $V$ is the unit cell volume and $\bm{v}_{q} = \frac{\partial \omega_{q}}{\partial \bm{q}}$ is the group velocity.
The vector $\bm{f}_{q}$ is a solution of the following linear equation:
\begin{align}
&\beta^{-1}\bm{v}_{q} \left( \frac{\partial n_{q}}{\partial T}\right) 
= \sum_{q'} (\bm{f}_{q} - \bm{f}_{q'}) \Lambda_{q}^{q'} \notag \\
& + \sum_{q',q''} \bigg[ (\bm{f}_{q} + \bm{f}_{q'} - \bm{f}_{q''})\Lambda_{qq'}^{q''} 
+ \frac{1}{2} (\bm{f}_{q} - \bm{f}_{q'} - \bm{f}_{q''})\Lambda_{q}^{q'q''} \bigg].
\label{eq:BTE_canonical2}
\end{align}
Here, $\Lambda_{q}^{q'}$ and $\Lambda_{qq'}^{q''} (\Lambda_{q}^{q'q''})$ are the intrinsic transition rates of phonon $q$ 
involving two and three phonons, respectively. 
For example, $\Lambda_{q}^{q'}$ includes the effect of phonon-isotope scattering and $\Lambda_{qq'}^{q''} (\Lambda_{q}^{q'q''})$
is dominated by three-phonon scattering processes associated with the cubic lattice anharmonicity.
To obtain the LTC of bulk solids, Eqs. (\ref{eq:LTC}) and (\ref{eq:BTE_canonical2}) must be solved on a fine $q$ mesh in the first BZ.
Broido \textit{et al.} first solved the linearized BTE for bulk silicon and germanium based on DFT and 
obtained $\kappa$ values that agree well with experimental data.\cite{2007ApPhL..91w1922B} 
Owing to the increased computational power and the development of efficient implementations,
it is now feasible to solve Eq.~(\ref{eq:BTE_canonical2}) for relatively complex materials 
using iterative algorithms,\cite{2007ApPhL..91w1922B,Fugallo:2013jl,2010PhRvB..82m4301C}
or even by using the direct method for simple systems.\cite{Chaput:2013ik,PhysRevB.91.094306}

To make the computation more feasible, the relaxation-time approximation (RTA) is often employed in the literature, whereby 
the LTC is simplified to
\begin{equation}
 \kappa_{\mathrm{RTA}} = \frac{\hbar^{2}}{NV k_{\mathrm{B}}T^{2}} \sum_{q} \omega_{q}^{2} \bm{v}_{q}\otimes\bm{v}_{q} n_{q}(n_{q} + 1) \tau_{q}. \label{eq:LTC_RTA}
\end{equation}
In the RTA, the current vertex corrections are neglected and the transport relaxation time is 
approximated by the quasiparticle lifetime $\tau_{q}$.\cite{2010PhRvB..82v4305S}
Therefore, the RTA incorrectly treats the normal process of three-phonon scattering as resistive, 
which results in $\kappa_{\mathrm{RTA}}<\kappa$.
The underestimation by the RTA is predominant in high-LTC materials such as diamond~\cite{PhysRevB.80.125203} as well as in a variety of two-dimensional materials.\cite{PhysRevB.82.115427,Cepellotti:2015ke,Jain:2015by,Zeraati:2016in}
Except for these cases, the RTA usually yields results similar to the full solution to the BTE.
Hence, the RTA has successfully been employed to analyze and predict the LTC of various kinds of solids.\cite{2011PhRvB..84h5204E,2011PhRvB..84j4302S,Tian2012,Li:2014dd,PhysRevLett.114.095501,Carrete:2014bz,PhysRevB.91.094306,Dekura:2013kv,PhysRevLett.115.205901}

While the predictive accuracy of the conventional BTE [Eq.~(\ref{eq:LTC})] and BTE-RTA [Eq.~(\ref{eq:LTC_RTA})] is reasonably high,
it has a limitation for severely anharmonic systems because the method treats the anharmonic effect perturbatively.
Most importantly, the conventional approach neglects the temperature dependence of phonon frequencies and eigenvectors.
This treatment prohibits us from estimating the LTC of high-temperature phases because of the imaginary modes within the HA.
To overcome this limitation, we made the following extension:\cite{2015PhRvB..92e4301T}
\begin{equation}
 \tilde{\kappa}_{\mathrm{RTA}} = \frac{\hbar^{2}}{NV k_{\mathrm{B}}T^{2}} \sum_{q} \Omega_{q}^{2} \tilde{\bm{v}}_{q}\otimes\tilde{\bm{v}}_{q} \tilde{n}_{q}(\tilde{n}_{q} + 1) \tilde{\tau}_{q}, \label{eq:LTC_SCP}
\end{equation}
where $\Omega_{q}$ is the SCP frequency, $\tilde{\bm{v}}_{q}=\frac{\partial \Omega_{q}}{\partial \bm{q}}$ is the group velocity of the renormalized phonon, and $\tilde{n}_{q}=n(\Omega_{q})$. The lifetime of the renormalized phonon $\tilde{\tau}_{q}$ associated with
the three-phonon scattering processes can be calculated from the diagram shown in Fig.~\ref{fig:SCP}(b).
This is similar to the original bubble diagram [Fig.~\ref{fig:diagrams} (b) and Eq.~(\ref{eq:bubble})], but the harmonic quantities
$\{\omega_{q},\bm{e}_{q}\}$ are replaced with the SCP solution $\{\Omega_{q},\bm{\epsilon}_{q}\}$.
We have successfully applied the new method to predict the LTC of cubic SrTiO$_{3}$ including its unusual temperature dependence,
which will be discussed in detail in Sect.~\ref{sec:appl}.
Other previous studies employed a similar approach for studying phonon transport properties in Bi$_{2}$Te$_{3}$,\cite{2014PhRvB..90m4309H} PbTe,\cite{Romero:2015km} and SrTiO$_{3}$,\cite{Feng:2015fc} in which 
the \textit{effective} harmonic and cubic IFCs were obtained by the temperature-dependent effective potential method.\cite{2011PhRvB..84r0301H,Hellman:2013ef}
These studies also indicated the importance of the renormalization of IFCs in the LTC of the studied materials.

Incorporating the four-phonon scattering process $\Lambda_{q}^{q'q''q'''}$ into the BTE [Eq.~(\ref{eq:BTE_canonical2})] is 
still a major computational challenge even within the RTA. 
In order to achieve this, the higher-order diagram shown in Fig.~\ref{fig:diagrams}(d) must be calculated with a dense $\bm{q}$ mesh, 
which has only been reported for the group IV semiconductors based on empirical potentials.~\cite{Feng:2016ib}
More work is thus needed to develop a robust understanding of the four-phonon scattering effects on the LTC of various kinds of solids including
low-dimensional systems.

MD is another powerful tool for calculating the LTC of solids. 
Since the anharmonicity is fully included, the MD-based methods, such as the non-equilibrium MD\cite{Evans:1982hd,1997JChPh.106.6082M} and Green-Kubo methods,\cite{Kubo:1957di}
are in principle more accurate than the BTE at high temperatures if accurate AIMD can be performed.
Moreover, they are suitable for systems with static and dynamical disorders, nanostructured devices, and aperiodic structures.
To achieve the reliable prediction of the LTC with MD, it is necessary to employ a reasonably large supercell to sample long-wavelength phonons
that make dominant contributions to the LTC.~\cite{2002PhRvB..65n4306S,He:2012hg,Tadano2014}
In addition, the simulation length must be sufficiently long to reduce statistical errors.
These requirements make the MD methods very costly, especially for high-LTC materials due to large mean free paths and long relaxation times of phonons. Therefore, straightforward applications of AIMD have only been attempted  for
a few materials having low LTC.\cite{2010PhRvL.104t8501S,*Stackhouse:2015ew,Yue:2016ew}
To make the computation feasible, classical force fields are commonly employed, but they can be a source of inaccuracy.
To address this problem, much effort has been made to improve the accuracy of classical force fields or to develop 
a new force field by using DFT results as training data.\cite{2010PhRvB..81t5441L,doi:10.1021/acs.nanolett.6b00457,2007PhRvL..98n6401B,2010PhRvL.104m6403B}

\subsection{First-principles calculation of force constants}
\label{sec:IFC}

To conduct the SCP calculations described in this section,
harmonic and quartic IFCs are necessary as inputs. 
Cubic terms are also required for performing the BTE calculations.
For that purpose, two conceptually different methods are commonly employed: density-functional perturbation theory (DFPT) and
the supercell method.

DFPT is one of the standard \textit{ab initio} methods for calculating
harmonic IFCs.\cite{1991PhRvB..43.7231G,1997PhRvB..5510337G,RevModPhys.73.515}
In DFPT, harmonic dynamical matrix and linear electron-phonon coupling coefficients for an arbitrary $\bm{q}$-point can be obtained.
Recently, an efficient implementation based on DFPT has also been developed for cubic IFCs.\cite{Paulatto:2013fb}
While DFPT is very accurate and efficient since it does not involve expensive supercell calculations,
its implementation is relatively complicated and its extension to quartic and higher-order IFCs is still challenging.

The supercell method is another approach to calculate IFCs, 
which is more expensive than DFPT but easier to implement.
In this approach, harmonic IFCs are estimated from the first-order numerical derivative of the atomic forces as
\begin{align}
 \Phi_{\mu\nu}(\ell\kappa;\ell'\kappa') = \frac{\partial^{2} U}{\partial u_{\mu}(\ell\kappa)\partial u_{\nu}(\ell'\kappa')}
  \approx -\frac{F_{\nu}[\ell'\kappa';u_{\mu}(\ell\kappa)]}{u_{\mu}(\ell\kappa)},
\end{align}
or inversely as\cite{1997PhRvL..78.4063P}
\begin{equation}
F_{\nu}[\ell'\kappa';u_{\mu}(\ell\kappa)] = - \Phi_{\mu\nu}(\ell\kappa;\ell'\kappa')u_{\mu}(\ell\kappa), \label{eq:fd2}
\end{equation}
which are valid when the displacement $u_{\mu}(\ell\kappa)$ is small, say $u_{\mu}(\ell\kappa) \sim$ 0.01 \AA.
$F_{\nu}[\ell'\kappa';u_{\mu}(\ell\kappa)]$ is the atomic force acting on the $\kappa'$th atom in the $\ell'$th cell
along direction $\nu$ when the displacement of $u_{\mu}(\ell\kappa)$ is made with all other atoms kept at the equilibrium positions.
Equation (\ref{eq:fd2}) is defined for every single coefficient $\Phi_{\mu\nu}(\ell\kappa;\ell'\kappa')$, many of which are
linearly dependent on each other because of the space group symmetry, the permutation symmetry, and the lattice periodicity.\cite{Esfarjani2008,Tadano2014}
For the convenience of the following discussion, let us introduce the column vectors $\bm{u}$ and $\bm{F}$, 
which consist of $3s$ atomic displacements and forces in the supercell containing $s$ atoms, respectively.
We also denote the $i$th displacement pattern as $\bm{u}_{i}$ and the associated force vector as $\bm{F}_{i}$.
Then, denoting a column vector comprising $n$ unique harmonic coefficients by $\bm{\Phi}_{2}$,
the set of linear equations [Eq.~(\ref{eq:fd2})] can be symbolically written as 
\begin{equation}
   \mathbb{A}_{2}\bm{\Phi}_{2} = \bm{\mathcal{F}}.
   \label{eq:A2}
\end{equation}
Here, $\mathbb{A}_{2}$ is a $3sm\times n$ matrix defined as $\mathbb{A}_{2}^{T} = [A_{2}^{T}(\bm{u}_{1}),\dots,A_{2}^{T}(\bm{u}_{m})]$, 
$A_{2}(\bm{u})=-\partial^{2} U_{2}/\partial \bm{\Phi}_{2}^{T}\partial \bm{u}$ is a $3s\times n$ matrix, 
$\bm{\mathcal{F}}^{T} = [\bm{F}_{1}^{T},\dots,\bm{F}_{m}^{T}]$, and $m$ is the number of displacement patterns.
Note that Eq.~(\ref{eq:A2}) is valid if the displacements are sufficiently small to neglect anharmonic effects in $\bm{\mathcal{F}}$.
If we consider sufficient displacement patterns and make the matrix $\mathbb{A}_{2}$ full rank, the harmonic IFCs can be
obtained, in the least-squares sense, as\cite{1997PhRvL..78.4063P,Chaput:2011dw}
\begin{equation}
   \bm{\Phi}_{2} = \mathbb{A}^{+}_{2} \bm{\mathcal{F}},
 \end{equation} 
 where $\mathbb{A}^{+}_{2}$ is the pseudoinverse of the matrix $\mathbb{A}_{2}$, which can be readily obtained by singular value decomposition.
A similar equation can be formed for the cubic terms as\cite{Chaput:2011dw} 
\begin{equation}
  \mathbb{A}_{3}\bm{\Phi}_{3} = \Delta\bm{\mathcal{F}}_{3},
  \label{eq:A3}
\end{equation}
in which $\Delta\bm{\mathcal{F}}_{3}= \bm{\mathcal{F}} - \mathbb{A}_{2}\bm{\Phi}_{2}$ is the cubic contribution to the atomic forces
and $\mathbb{A}_{3}^{T} = [A_{3}^{T}(\bm{u}_{1}),\dots,A_{3}^{T}(\bm{u}_{m})]$ with $A_{3}(\bm{u})$ defined as
$A_{3}(\bm{u})=-\partial^{2} U_{3}/\partial \bm{\Phi}_{3}^{T}\partial \bm{u}$.
Again, Eq.~(\ref{eq:A3}) assumes that the quartic and higher-order contributions to atomic forces $\bm{\mathcal{F}}$ are negligible,
which is valid when the displacement is small.
To make the matrix $\mathbb{A}_{3}$ full rank, we need to consider many displacement patterns where 
at least two atoms must be displaced simultaneously.
Therefore, their calculation is more expensive than the harmonic terms. 
To reduce the computational burden, it is common to introduce a cutoff distance and/or 
employ a smaller supercell for cubic IFCs.
These treatments are well justified if the anharmonic interaction is short-range, which is
usually the case in covalent compounds. 
By contrast, long-range cubic interactions may be significant in ionic compounds,\cite{Li:2015gf}
for which a careful choice of the cutoff radius must be made to avoid erroneous numerical results. 

Following the above procedure, it is straightforward to form linear equations for quartic and 
higher-order IFCs, but they are difficult to solve because of a large number of DFT calculations
necessary to make the matrix $\mathbb{A}_{n} (n\geq 4)$ full rank. 
Moreover, an accurate DFT calculation of $\Delta\bm{\mathcal{F}}_{n}$ is difficult because 
each $\Delta\bm{\mathcal{F}}_{n}$ is very small in the small-displacement limit. 
While this numerical issue may be avoided by using a larger displacement, 
finding an appropriate value for the displacement is formidable as it depends on the atomic environment.
Performing AIMD simulation is a reasonable solution to this problem as it can generate physically relevant atomic configurations automatically. 
From the displacement-force data set obtained from the AIMD trajectory, 
we may estimate IFCs by solving the ordinary least-squares (OLS) problem\cite{Tadano2014}
\begin{equation}
  \tilde{\bm{\Phi}} = \argmin_{\bm{\Phi}} \| \mathbb{A}\bm{\Phi} - \bm{\mathcal{F}} \|^{2}_{2}.
  \label{eq:OSL}
\end{equation}
Here, $\bm{\Phi}$ is the column vector defined as $\bm{\Phi}^{T} = [\bm{\Phi}_{2}^{T}, \bm{\Phi}_{3}^{T}, \bm{\Phi}_{4}^{T}, \dots]$ and
$\mathbb{A}^{T} = [A^{T}(\bm{u}_{1}),\dots,A^{T}(\bm{u}_{m})]$ with 
$A(\bm{u})=-\partial^{2} U/\partial \bm{\Phi}^{T}\partial \bm{u}$. 
Given a sufficient number of AIMD snapshots, the unknown IFCs can be estimated as
$\tilde{\bm{\Phi}} = \mathbb{A}^{+}\bm{\mathcal{F}}$. 
Using this approach, we successfully extracted anharmonic IFCs of Si and Mg$_{2}$Si from AIMD trajectories.\cite{Tadano2014}
Recently, Zhou \textit{et al.} proposed a more sophisticated approach called the compressed sensing lattice dynamics (CSLD) method,\cite{PhysRevLett.113.185501}
where they estimated IFCs using the least absolute shrinkage and selection operator (LASSO)\cite{Tibshirani:1996wb}:
\begin{equation}
    \tilde{\bm{\Phi}} = \argmin_{\bm{\Phi}} \| \mathbb{A}\bm{\Phi} - \bm{\mathcal{F}} \|^{2}_{2} + \lambda \|\bm{\Phi}\|_{1}.
    \label{eq:LASSO}
\end{equation} 
Owing to the $\ell_{1}$-regularization term, physically irrelevant IFCs are driven to be exactly zero and important terms
are selected and calculated automatically. The method is robust against random and systematic noise\cite{CPA:CPA20124} and 
a penalty term helps to avoid the overfitting inherent in the OLS method.
The coefficient $\lambda$ is a hyperparameter that controls the
trade-off between the sparsity and accuracy of the model; all IFCs will be zero in the large-$\lambda$ limit and $\lambda=0$ corresponds to the OLS regression.
An appropriate value of $\lambda$ can be estimated, for instance, by cross-validation, as will be demonstrated in Sect.~\ref{sec:appl}.

In Ref.~\onlinecite{PhysRevLett.113.185501}, Zhou \textit{et al.} extracted harmonic and anharmonic IFCs of Si, NaCl, and Cu$_{12}$Sb$_{4}$S$_{13}$
using Eq.~(\ref{eq:LASSO}) combined with AIMD simulation. On the basis of the IFCs estimated by LASSO, they calculated the LTC of these materials and
obtained good agreement with experimental data. Using the same approach, we successfully extracted anharmonic IFCs of 
SrTiO$_{3}$,\cite{2015PhRvB..92e4301T} superconducting hydrogen sulfides,\cite{2016PhRvB..93i4525S} and 
thermoelectric clathrates and applied them to SCP and BTE calculations. 
Also, compressed sensing has successfully been employed to develop interatomic potentials\cite{Seko:2014ke,*Seko:2015gq} 
and a cluster expansion model\cite{2013PhRvB..87c5125N} based on DFT.

\section{Applications to Strontium Titanate}
\label{sec:appl}

In Ref.~\onlinecite{2015PhRvB..92e4301T}, we calculated anharmonic phonon properties of cubic SrTiO$_{3}$ (c-STO), which is
stable above 105 K. 
All of the DFT calculations were performed using the Vienna Ab initio Simulation Package (\textsc{vasp}),\cite{Kresse1996,Kresse1999}
where the generalized gradient approximation for solids\cite{PBEsol} (PBEsol) 
was employed for the exchange correlation functional. 
This choice was made because the PBEsol functional is known to predict the lattice constants
for a wide class of materials, which is crucial for the accurate calculation of phonon properties.\cite{Kresse_functional,Skelton:2014db}
Using quartic IFCs estimated by LASSO, we performed SCP simulations and obtained overall good agreement with
experimentally observed phonon spectra at room temperature. 
However, the quantitative accuracy based on the PBEsol functional was not very satisfactory because 
the calculated frequencies of the ferroelectric (antiferrodistortive) soft modes were overestimated (underestimated).

In this paper, we report more accurate computational result within 
the Heyd-Scuseria-Ernzerhof hybrid functional (HSE06, hereafter HSE)\cite{HSE,*Heyd:2006dc}.
Before performing phonon calculations, we optimized the lattice constant of c-STO using the HSE functional and obtained 3.900 \AA,  
in good agreements with the experimental value of 3.905 \AA{} (Ref.~\onlinecite{Okazaki1973545}, 293 K) 
and the previous HSE result of 3.904 \AA.\cite{Kresse_functional}
The computational conditions employed in this study were basically the same as those in the previous PBEsol study,\cite{2015PhRvB..92e4301T} 
except that BZ integration was performed with the 8$\times$8$\times$8 Monkhorst-Pack $\bm{k}$-point grid.
All of the phonon calculations reported here were performed with a 2$\times$2$\times$2 cubic supercell (40 atoms) using the \textsc{alamode} package,\cite{alamode,Tadano2014} an open-source software
developed for modeling lattice anharmonicity and thermal conductivity.

\subsection{Anharmonic force constants by LASSO}
\label{sec:ifc_lasso}

To develop an accurate model based on Eq.~(\ref{eq:Taylor}), we considered anharmonic terms up to the sixth order.
Here, all possible cubic terms occurring in the 2$\times$2$\times$2 supercell were considered.
For quartic IFCs, we only considered onsite, two-body, and three-body IFCs within a cutoff length of 6 \AA\ and neglected
less important four-body terms. Also, only onsite and two-body terms within the same cutoff radius were considered 
for the fifth- and sixth-order IFCs. 
We then determined a set of linearly independent parameters by making full use of the available symmetry operations 
and the constraints for the translational invariance.\cite{Esfarjani2008,Tadano2014}
The numbers of independent parameters were found to be 698, 2215, 43, and 125 for cubic, quartic, quintic, 
and sextic terms, respectively, which in total form a parameter vector $\bm{\Phi}$ of length 3018.
Calculating all elements of $\bm{\Phi}$ would be unfeasible if the conventional finite displacement
method were employed as it would need about 5,000 DFT calculations.
To determined these IFCs by LASSO, we prepared 40 displacement-force data sets using AIMD, 
whose detailed procedure is described in Ref.~\onlinecite{2015PhRvB..92e4301T}.
After that, the data set was split into four smaller subsets and the hyperparameter $\lambda$ in 
Eq.~(\ref{eq:LASSO}) was selected by four-fold cross-validation.\cite{Hastie:2009fg}
The solution path was obtained by using the coordinate descent method,\cite{Hastie:2015tb}
for which each column vector of the matrix $\mathbb{A}$ was standardized beforehand.
Upon solving Eq.~(\ref{eq:LASSO}), harmonic IFCs were fixed to the values obtained by the OLS method.

\begin{figure}
\centering
\includegraphics[width=8.5cm,clip]{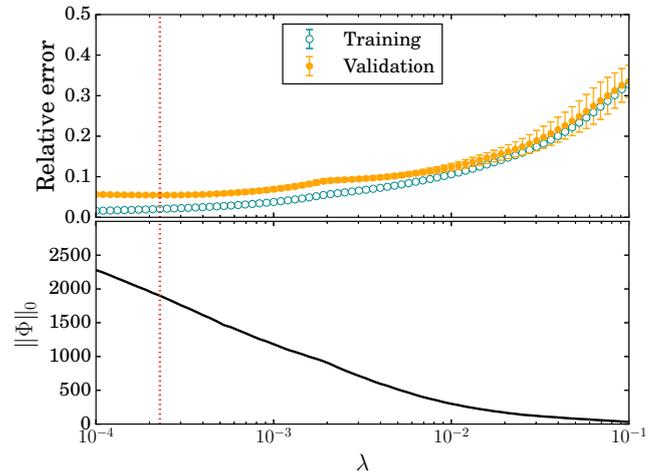}
\caption{(Color online) (a) Relative errors in atomic forces and (b) number of nonzero coefficients 
as a function of the hyperparameter $\lambda$ obtained for PBEsol. 
The dotted vertical line indicates the selected value of $\lambda$ that minimizes the cross-validation score.}
\label{fig:cv}
\end{figure}

Figure \ref{fig:cv} shows the result of the cross-validation obtained for PBEsol. 
Since very similar results were obtained for HSE, they are not shown here.
The top panel shows the relative error of the atomic forces, which is
defined as the square root of $\| \mathbb{A}\tilde{\bm{\Phi}} - \bm{\mathcal{F}} \|^{2}_{2}/\| \bm{\mathcal{F}}\|_{2}^{2}$.
In the large -$\lambda$ region, the difference between the training error and the validation error is marginal.
With decreasing $\lambda$, the difference becomes more predominant; 
the training error monotonically decreases, whereas the validation curve has a minimum around 
$\lambda=2.3\times 10^{-4}$, indicated by the dotted vertical line in the figure. 
This value of $\lambda$ was selected as an optimal choice because it is expected to give
the best prediction accuracy for independent data sets.
The bottom panel shows the number of nonzero coefficients. 
With the optimal value of $\lambda$, we obtained 2024 nonzero coefficients, which is about
67\% of the total number of elements. 
The accuracy of the extracted IFCs $\tilde{\bm{\Phi}}$ was carefully verified by applying them
to the independent atomic configurations sampled by AIMD.

\subsection{SCP solution with HSE}

Using the calculated harmonic and quartic IFCs, we conducted SCP calculations based on Eqs.~(\ref{eq:SCP1}) and (\ref{eq:SCP2}).
To correctly describe the PM, we considered off-diagonal elements of the loop diagram as in Ref.~\onlinecite{2015PhRvB..92e4301T}.
The convergence of the anharmonic frequencies $\{\Omega_{q}\}$ at the commensurate 2$\times$2$\times$2 $\bm{q}$-point grid was
carefully checked with respect to the intermediate $\bm{q}_{1}$ grid in Eq.~(\ref{eq:SCP2}) and found sufficiently converged
results with 8$\times$8$\times$8 $\bm{q}_{1}$ points. 
In addition, the non-analytic correction to the harmonic dynamical matrix was included to reproduce the LO-TO splitting around the $\Gamma$ point. To this end, we employed the Born effective charges and the dielectric constant reported in Ref.~\onlinecite{2015PhRvB..92e4301T}
for PBEsol. Calculating these quantities using the HSE functional was very expensive. Therefore, as an approximation, we employed the Born effective charges obtained by PBEsol and the experimental dielectric constant of 5.18\cite{1994PhRvL..72.3618Z} in the following HSE calculations.

\begin{figure*}
\centering
\includegraphics[width=0.9\textwidth]{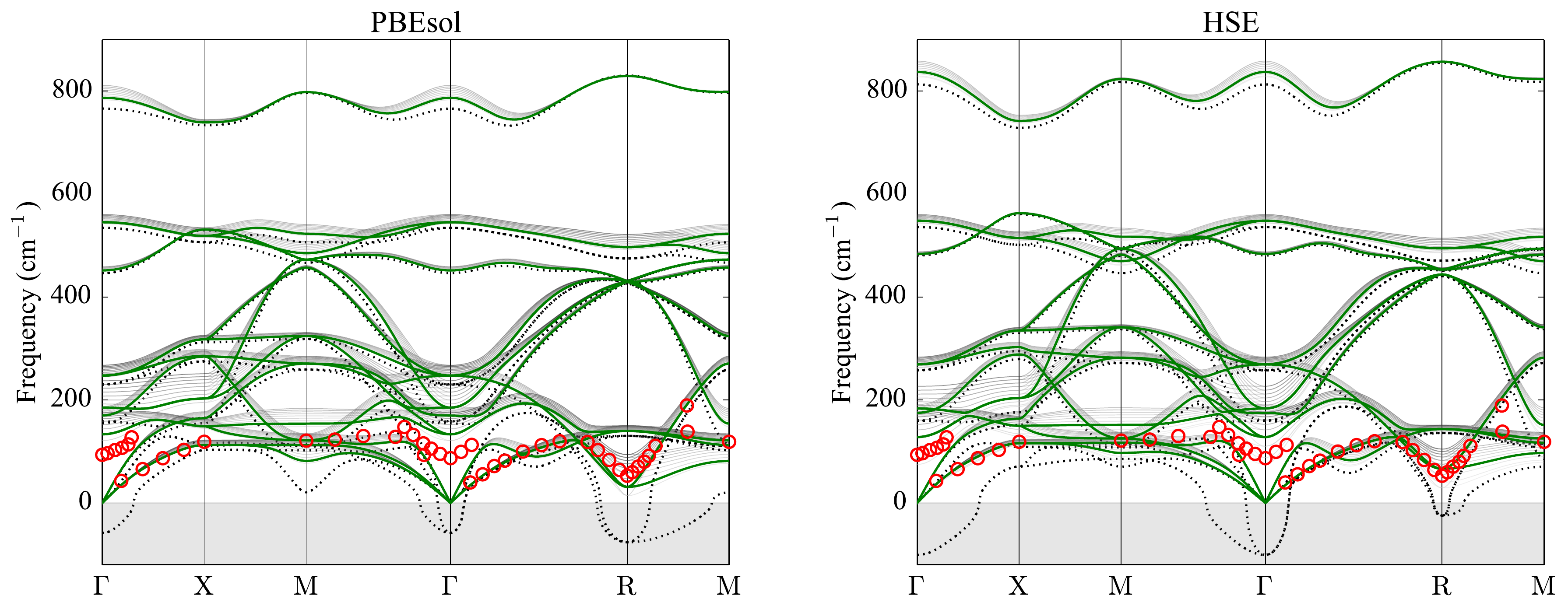}
\caption{(Color online) Temperature-dependent anharmonic phonon dispersion of cubic SrTiO$_{3}$ calculated with the PBEsol (left) and HSE (right) exchange-correlation functionals. The gray solid lines in the left (right) figure represent the SCP solutions at different temperatures ranging from 200 (0) to 1000 K. 
The dotted lines are harmonic lattice dynamics results and the red open circles are experimental values at 300 K adapted from Refs.~\onlinecite{Stirling1972} and \onlinecite{Cowley1969181}. Theoretical curves at 300 K are highlighted as thick solid lines for comparison with the experimental data.}
\label{fig:SCP_bands}
\end{figure*}

Figure \ref{fig:SCP_bands} shows the phonon dispersion curves of c-STO calculated within the HA (dotted lines) 
and the SCP theory (thin solid lines). For the SCP results, we show several curves corresponding to different temperatures 
ranging from 200 (0) to 1000 K for PBEsol (HSE) in steps of 100 K. For ease of the comparison with 
inelastic neutron scattering (INS) data measured at 300 K,\cite{Stirling1972,Cowley1969181}  the theoretical curves at 300 K are highlighted as thick solid lines. 
In the harmonic phonon dispersion, unstable phonon modes ($\omega_{q}^{2} < 0$) occur at points $\Gamma\;(0,0,0)$ and 
$\mathrm{R}\;(\frac{1}{2},\frac{1}{2},\frac{1}{2})$, which correspond to the ferroelectric (FE) and 
antiferrodistortive (AFD) modes, respectively.
The harmonic phonon frequency of the FE mode is 101$i$ cm$^{-1}$ with HSE, whose absolute value is larger than the
PBEsol result of 58$i$ cm$^{-1}$, indicating the enhanced instability of the FE mode by the Fock exchange term.
For the AFD mode, the trend was opposite and the frequency changed from 76$i$ cm$^{-1}$ (PBEsol) to 25$i$ cm$^{-1}$ (HSE) as 
summarized in Table \ref{table:frequency}.
These tendencies are consistent with the previous DFT study of Wahl \textit{et al.}\cite{Kresse_functional}

As can be seen in Fig.~\ref{fig:SCP_bands}, the quartic anharmonicity generally increases the vibrational frequencies in c-STO, 
which is particularly predominant in the low-energy optical modes. 
The PBEsol result obtained here is almost identical to our previous result in Ref.~\onlinecite{2015PhRvB..92e4301T}.
The SCP frequency of the FE mode $\Omega_{\mathrm{FE}}$ is 133 cm$^{-1}$ at 300 K with PBEsol, which overestimates the INS data of 87.1$\pm$5.6 cm$^{-1}$ (Ref.~\onlinecite{doi:10.1143/JPSJ.26.396}, 293 K) and the IR data of 89 cm$^{-1}$ (Ref.~\onlinecite{PhysRevB.22.5501}, 300 K).
By contrast, the frequency of the AFD mode at point R is underestimated with PBEsol.
We have found that the underestimation at point R can be reduced by employing HSE 
as shown in the right panel of Fig.~\ref{fig:SCP_bands}.
However, $\Omega_{\mathrm{FE}}$ was improved only slightly by HSE and the SCP frequency of 128 cm$^{-1}$ 
still overestimates the experimental values.
In what follows, we discuss the origin of the disagreement by going beyond the SCP theory.

\begin{table*}
\centering
\caption{\label{table:frequency} Phonon frequencies (cm$^{-1}$) of the $\Gamma_{15}$ (FE) and $R_{25}$ (AFD) soft modes in c-STO calculated with different levels of approximation.
Calculated anharmonic frequencies are evaluated at 300 K.}

\begin{ruledtabular}
\begin{threeparttable}
\begin{tabular}{ccccc p{1pt} ccccc} 
 &\multicolumn{4}{c}{PBEsol} & & \multicolumn{4}{c}{HSE} &  \multirow{2}{*}{Expt.}\\  \cline{2-5} \cline{7-10} 
 &HA & SCP & SCP + B &  Peak in $A_{q}(\omega)$ & &HA & SCP& SCP + B & Peak in $A_{q}(\omega)$ & \\ \hline
$\Gamma_{15}$ (FE)&58$i$~\tnote{a)} & 133 & 120 & 120 & &101$i$ & 128 & 109 & 104 & 87.1$\pm$5.6~\tnote{b)},~ 89~\tnote{c)} \\ 
$R_{25}$ (AFD) &76$i$~\tnote{a)} &  31 & 2&  18& & 25$i$ & 66 &  53&  52& 48.7$\pm$1.8~\tnote{d)} \\ 
\end{tabular}
\begin{tablenotes}\footnotesize
\item[a] Ref.~\onlinecite{2015PhRvB..92e4301T}
\item[b] Ref.~\onlinecite{doi:10.1143/JPSJ.26.396} (INS, 293 K)
\item[c] Ref.~\onlinecite{PhysRevB.22.5501} (IR, 300 K)
\item[d] Ref.~\onlinecite{Cowley1969181} (INS, 300 K)
\end{tablenotes}
\end{threeparttable}
\end{ruledtabular}
\end{table*}

\subsection{Frequency shift by cubic anharmonicity}

Here, we discuss the effect of the cubic anharmonicity on phonon frequencies. 
In the present SCP calculations, only the quartic anharmonicity is included as diagrammatically shown in Fig.~\ref{fig:SCP}(a).
However, frequency shifts due to the cubic terms may not be negligible. 
To reveal its effect quantitatively, we calculated the bubble self-energy shown in Fig.~\ref{fig:SCP}(b) and 
estimated the frequency shift of the FE mode from
 $\Delta_{q}^{\mathrm{bubble}}=-\frac{1}{\hbar}\mathrm{Re}\tilde{\Sigma}_{q}^{\mathrm{bubble}}(\Omega_{q})$.
The results of $\Omega_{q} + \Delta_{q}^{\mathrm{bubble}}$ are shown in Table \ref{table:frequency} as ``SCP+B'' (B stands for bubble).
The correction $\Delta_{q}^{\mathrm{bubble}}$ was found to be negative and significant for both of the FE and AFD modes at 300 K.
For the FE mode, it was about $-13$ cm$^{-1}$ with PBEsol and $-19$ cm$^{-1}$ with HSE.
For the AFD mode, the correction was very large, $\Delta_{q}^{\mathrm{bubble}} \approx -29$ cm$^{-1}$ with PBEsol, and
the corrected frequency $\Omega_{q} + \Delta_{q}^{\mathrm{bubble}}$ was almost zero.
It is interesting to note that the frequency dependence of $\mathrm{Re}\tilde{\Sigma}_{q}^{\mathrm{bubble}}(\omega)$ was
marginal in the low-frequency region ($\omega < 200$ cm$^{-1}$) and the difference between $\Delta_{q}^{\mathrm{bubble}}$ and
$-\frac{1}{\hbar}\mathrm{Re}\tilde{\Sigma}_{q}^{\mathrm{bubble}}(0)$ was minor, especially when the SCP frequency was small.
Therefore, it may be reasonable to make the static approximation of $\tilde{\Sigma}_{q}^{\mathrm{bubble}}(\omega)\approx\tilde{\Sigma}_{q}^{\mathrm{bubble}}(0)$ in order to calculate the frequency shift due to the bubble diagram, as carried out in Ref.~\onlinecite{1998PhRvB..58.2596S}.
With the correction from the bubble diagram, the theoretical values based on HSE are in better agreement with the experimental values.
Although other theoretical and numerical aspects must be carefully investigated, 
our results indicate the importance of accurate theoretical and computational modeling of anharmonic effects 
for robust understanding of the lattice dynamics in severely anharmonic materials.

\begin{figure}
\centering
\includegraphics[width=8.5cm, clip]{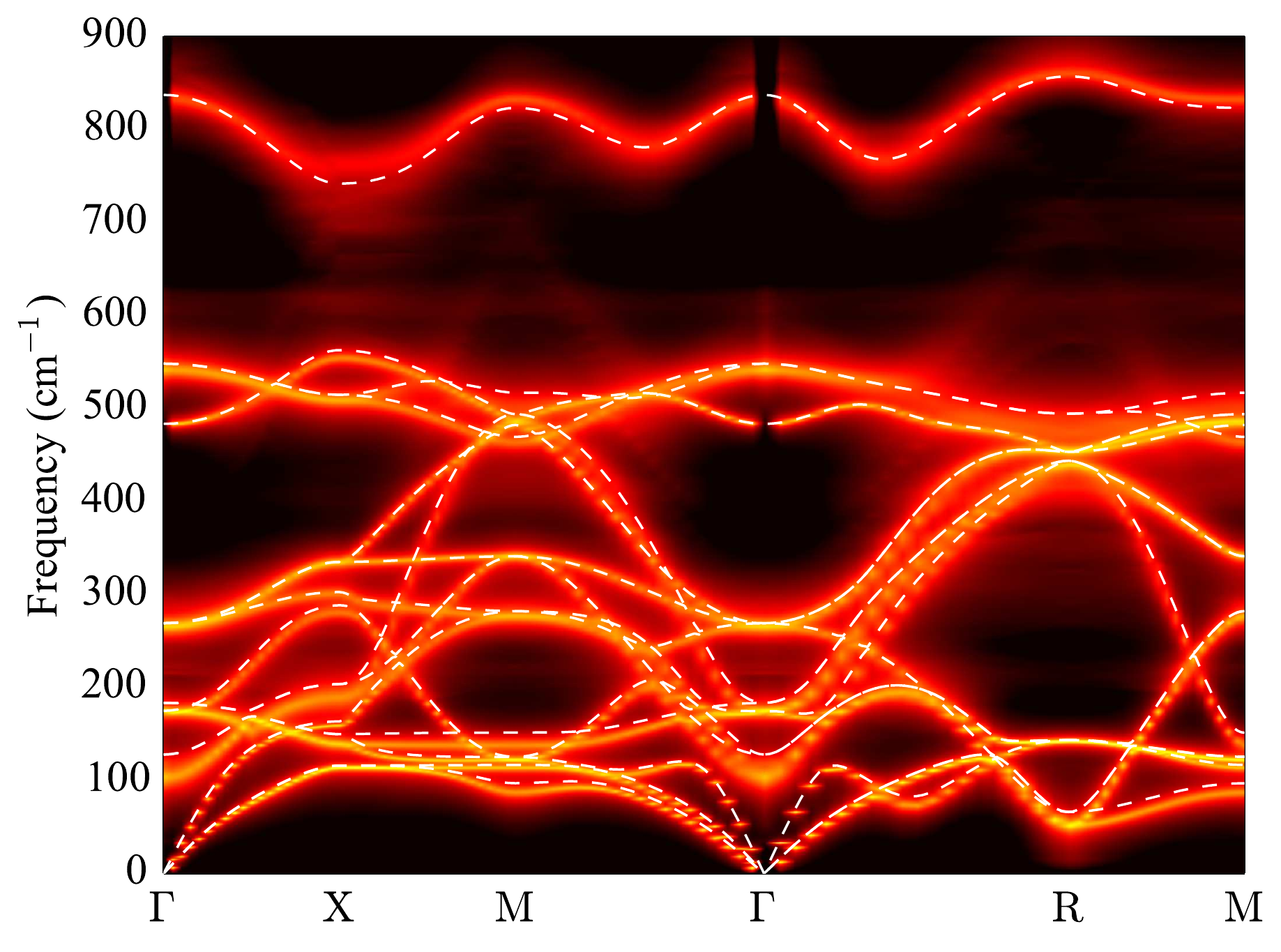}
\caption{(Color online) Phonon spectral function of cubic SrTiO$_{3}$ at 300 K calculated with the HSE functional. 
The SCP solution at 300 K is also shown by white dashed lines for comparison.}
\label{fig:spectral}
\end{figure}

To make a straightforward comparison with the experimental data, we also calculated the spectral function $A_{q}(\omega) = -\frac{1}{\pi}\mathrm{Im}G_{q}(\omega)$ using the following formula:
\begin{equation}
  A_{q}(\omega) \propto \frac{4\Omega_{q}^{2}\Gamma_{q}^{\mathrm{bubble}}(\omega)}{\{\omega^{2}-\Omega_{q}^{2}-2\Omega_{q}\Delta_{q}^{\mathrm{bubble}}(\omega)\}^{2}+\{2\Omega_{q}\Gamma_{q}^{\mathrm{bubble}}(\omega)\}^{2}}.
  \label{eq:spec}
\end{equation}
Since $A_{q}(\omega)$ has a direct connection with the dynamical structural factor measured in INS experiments,
it should be more reasonable and accurate to estimate theoretical phonon frequencies from peak positions of Eq.~(\ref{eq:spec}).
The estimated frequencies are also shown in Table \ref{table:frequency}, which differ from the ``SCP+B'' results. 
This indicates the limited accuracy of the SCP+B results when the condition of $\hbar\Omega_{q} \gg |\Sigma_{q}|$ is not well satisfied.
Figure~\ref{fig:spectral} shows the spectral function at 300 K calculated along the high-symmetry lines of the BZ.
For comparison, we also show the SCP dispersion curves by dashed lines.
It is evident from the figure that the AFD mode and high-energy optical modes above $\sim 400$ cm$^{-1}$ are strongly damped by 
the cubic anharmonicity. Moreover, the frequency shift caused by the bubble diagram is most significant in the low-lying optical modes
around point $\Gamma$ and along path R--M.

\subsection{Anomalous thermal transport in SrTiO$_{3}$}

\begin{figure}
\centering
\includegraphics[width=8.5cm,clip]{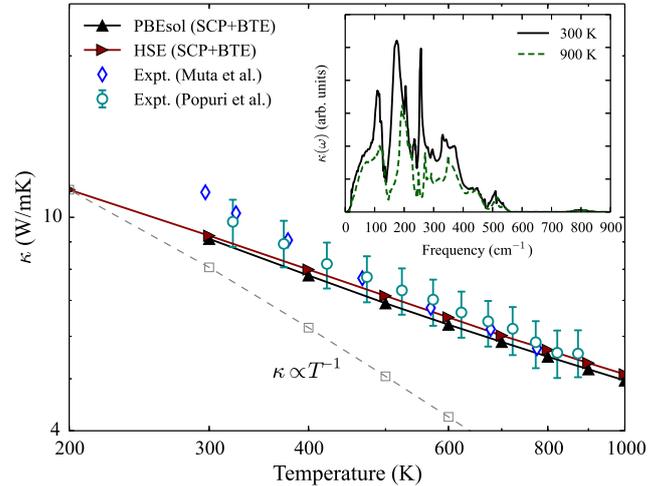}
\caption{(Color online) Temperature dependence of calculated LTC in cubic SrTiO$_{3}$ compared
with experimental values adapted from Refs.~\onlinecite{Muta2005306,C4RA06871H}.
The dashed line shows LTC values calculated without changing the SCP frequencies from those at 200 K,
which approximately follow $\kappa\propto T^{-1}$.  
(Inset) Thermal conductivity spectra at 300 K and 900 K obtained with the HSE functional.}
\label{fig:kappa}
\end{figure}

Finally, we calculated the LTC of c-STO using Eq.~(\ref{eq:LTC_SCP}) with 20$\times$20$\times$20 $\bm{q}$-grid points.
In Fig.~\ref{fig:kappa}, we compare our computational results with experimental data from Refs.~\onlinecite{Muta2005306,C4RA06871H}.
As can be seen in the figure, we obtained overall good agreement with the experimental LTC including its anomalous temperature dependence.
Unlike the case of soft-mode frequencies discussed in the previous section, 
the improvement made by the accurate HSE functional was minor. 
This is because the LTC is a quantity averaged over the first BZ, for which the prediction accuracy of 
the less expensive PBEsol functional should be sufficient.

The anomalous temperature dependence of the LTC ($\kappa\propto T^{-\alpha}$, $\alpha\sim$ 0.6--0.7) can be attributed to
the hardening of phonon frequencies caused by the quartic anharmonicity. 
When we employed the SCP frequencies at 200 K for BTE calculations at higher temperatures, 
we obtained the conventional temperature dependence of $\kappa\propto T^{-1}$ as shown by the dashed line in Fig.~\ref{fig:kappa}.
However, such a treatment cannot be justified because the LTC values are significantly underestimated at high temperatures,
which was also discussed in Refs.~\onlinecite{Romero:2015km,Stackhouse:2015ew,vanRoekeghem:2016kd}.
We think the effect of the frequency renormalization by the quartic terms can be significant in other ultralow-LTC 
materials such as SnSe and clathrate, which will be a topic of future work.

It is interesting to note that in c-STO more than 70\% of the total thermal conductivity is carried by 
high-frequency optical modes above 150 cm$^{-1}$,\cite{Feng:2015fc,2015PhRvB..92e4301T} as can be seen in the inset of Fig.~\ref{fig:kappa}. 
Here, $\kappa(\omega)$ is the thermal conductivity spectrum defined as 
$\kappa(\omega) = \sum_{q}\kappa_{q}\delta(\omega-\omega_{q})$ with $\kappa_{q}$ 
being the contribution from each phonon mode $q$.
This can mainly be attributed to the large group velocities of these optical modes, which are larger than
those of the acoustic modes.

\section{Summary and Conclusions}
\label{sec:conclusion}

We reviewed our recent development of a first-principles framework to study lattice dynamical properties of
strongly anharmonic crystals. By combining an efficient implementation based on the self-consistent phonon (SCP) theory and 
the compressed sensing of anharmonic force constants, phonon frequencies renormalized by the quartic anharmonicity 
can be calculated nonperturbatively at various temperatures.
In addition, the intrinsic phonon linewidths and frequency shifts by the cubic anharmonicity can be calculated by 
considering the bubble diagram on top of the SCP lattice-dynamics wavefunctions, which is essential for predicting
the lattice thermal conductivity (LTC) of anharmonic crystals from first principles.

We demonstrated the high predictive accuracy of the developed computational approach by carefully comparing soft-mode frequencies and 
LTC values of cubic SrTiO$_{3}$ with available experimental data, for which overall good agreement was obtained.
In addition, we showed that accurate, albeit expensive, calculation based on the HSE hybrid functional was feasible within
the present framework, where a marked improvement was achieved in the frequency of the antiferrodistortive soft mode.
We expect the presented approach to open up a wide range of applications and provide more insight into
anharmonic effects in severely anharmonic materials, such as ferroelectric, thermoelectric, and photovoltaic materials, 
as well as in light-element superconductors where the conventional harmonic approximation and the perturbative treatments break down.

\section*{Acknowledgements}

This study was partly supported by JSPS KAKENHI Grant Number 16K17724 and ``Materials research by Information Integration'' Initiative (MI2I) project
of the Support Program for Starting Up Innovation Hub from Japan Science and Technology Agency (JST).
The computation in this work was carried out using the facilities of the Supercomputer Center,
Institute for Solid State Physics, The University of Tokyo.

\end{document}